\documentclass[conference]{IEEEtran}
\IEEEoverridecommandlockouts
\usepackage{cite}
\usepackage{amsmath,amssymb,amsfonts}
\usepackage{algorithmic}
\usepackage{graphicx}
\usepackage{textcomp}
\usepackage{xcolor}
\usepackage{caption}
\usepackage{subcaption}
\usepackage{multirow}
\usepackage{orcidlink}
\def\BibTeX{{\rm B\kern-.05em{\sc i\kern-.025em b}\kern-.08em
    T\kern-.1667em\lower.7ex\hbox{E}\kern-.125emX}}
\begin{document}

% \title{Optimizing OpenAI Whisper for Aphasic Speech Recognition: Fine‑Tuning Strategies and Data‑Ratio Analysis\\}
\title{SHAP-AAD: DeepSHAP-Guided Channel Reduction for EEG Auditory Attention Detection}

\author{Rayan Salmi*, Guorui Lu*\orcidlink{0009-0005-3455-0670},~\IEEEmembership{Student Member,~IEEE},
Qinyu Chen\orcidlink{0000-0002-3284-4078},~\IEEEmembership{Member,~IEEE}
\thanks{*Rayan Salmi and Guorui Lu equally contribute to the work.}
\thanks{Rayan Salmi, Guorui Lu and Qinyu Chen are with the Leiden Institute of Advanced Computer Science (LIACS), Leiden University, The Netherlands. (q.chen@liacs.leidenuniv.nl).}
}

\maketitle

\begin{abstract}
Electroencephalography (EEG)-based auditory attention detection (AAD) offers a non-invasive way to enhance hearing aids, but conventional methods rely on too many electrodes, limiting wearability and comfort. This paper presents \textbf{SHAP-AAD}, a two-stage framework that combines DeepSHAP-based channel selection with a lightweight temporal convolutional network (TCN) for efficient AAD using fewer channels. DeepSHAP, an explainable AI technique, is applied to a Convolutional Neural Network (CNN) trained on topographic alpha-power maps to rank channel importance, and the top-$k$ EEG channels are used to train a compact TCN. Experiments on the DTU dataset show that using 32 channels yields comparable accuracy to the full 64-channel setup (79.21\% vs. 81.06\%) on average. In some cases, even 8 channels can deliver satisfactory accuracy. These results demonstrate the effectiveness of SHAP-AAD in reducing complexity while preserving high detection performance.

% These improvements demonstrate that fine-tuned AS-ASR provides higher accuracy and low latency and serves as a viable solution for real-time on-device pathological speech recognition.
\end{abstract}

\begin{IEEEkeywords}
AAD, EEG, channel reduction, Shapley value, explainable AI.
\end{IEEEkeywords}

\section{Introduction}

Humans are capable of selectively focusing on a specific sound source in a noisy environment, a phenomenon known as the cocktail party effect~\cite{haykin2005cocktail}, which plays a crucial role in daily life and communication. However, hundreds of millions of adults worldwide struggle with hearing loss~\cite{cunningham2017hearing}, which significantly impairs their ability to suppress competing sounds~\cite{dai2018sensorineural} and shows reduced capacity for selective auditory attention in noisy environments~\cite{ciccarelli2019comparison}. Therefore, it is important to develop methods that can help users focus on a desired sound source while filtering out distracting noise from other directions.

Recently, there have been advances in neuroscience that have shown that auditory attention can be detected from signals from brain activity through the electroencephalogram (EEG)~\cite{fan2025seeing, cai2022neural, lan2025low}. Due to the non-invasive nature, portability, and cost-effectiveness~\cite{Abiri_2019}, EEG is particularly well-suited for long-term and wearable use, making it align well with the usage scenarios of auditory attention detection (AAD).

Deep learning techniques have shown great potential in addressing EEG-based AAD. Convolutional neural networks (CNN)~\cite{vandecappelle2021eeg}, recurrent neural networks (RNN)~\cite{eskandarinasab2024gru}, attention mechanisms~\cite{yan2024darnet, cai2021eeg}, and spiking neural networks (SNN)~\cite{cai2022neural} have been explored for this task. However, these methods commonly use 64-channel EEG signals as input. Such a large number of channels leads to high hardware complexity in EEG acquisition devices, which in turn increases the cost and reduces comfort, thus limiting the applicability of these methods in long-term and wearable scenarios, precisely the contexts in which AAD is most relevant.

A few studies have considered scenarios with reduced channel counts, such as the group-LASSO method~\cite{narayanan2018effect}, the decoder magnitude-based method~\cite{mirkovic2015decoding}, and the greedy utility-based method~\cite{narayanan2019analysis}. In contrast to these approaches, in this study, we propose an explainable deep learning-based selection strategy to reduce the number of channels. This strategy treats each channel as a feature and applies DeepSHAP~\cite{chen2020explaining} to quantify the contribution of each channel to the output. Channels with low contributions are likely to be excluded to reduce the total number of channels.
To the best of our knowledge, this is the first channel selection method using explainable AI in AAD tasks, exploring a new possibility of reducing the number of channels in the relevant applications.

% \begin{figure}[htbp]
% \centerline{\includegraphics[width=\columnwidth]{Graph1.jpg}}
% \caption{Comparison between conventional and aphasia-adapted ASR systems on disfluent speech.}
% \label{fig:intro}
% \end{figure}

\section{Background}

The Shapley value~\cite{shapley1953value}, originating from cooperative game theory, provides a principled way to fairly distribute the total gain of a game among its players based on their individual contributions. The idea was then introduced to machine learning as an explainable AI method to analyze the contributions of features to the model by treating the model as the "game" and the input features as the "players"~\cite{merrick2020explanation}.

However, computing the exact Shapley values requires evaluating all $2^n$ possible subsets of features, which is infeasible for models such as neural networks. To solve this problem, DeepSHAP~\cite{chen2020explaining} is proposed. The method combines the theoretical foundation of SHAP and the backpropagation strategy of Deep Learning Important FeaTures (DeepLIFT)~\cite{shrikumar2017learning}, making it suitable for approximating feature contributions in complex models.

\section{Methodology}
% \subsection{Overall Explainability-Guided EEG Channel Reduction Pipeline}

\begin{figure*}[tbp]
\centerline{\includegraphics[width=2.1\columnwidth]{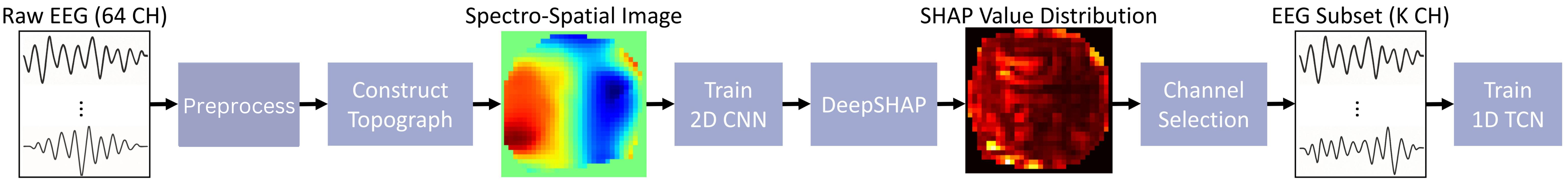}}
\caption{The overall workflow of the proposed SHAP-AAD framework.}
\label{fig:workflow}
\end{figure*}
The overall workflow of the proposed SHAP-AAD framework, shown in Fig.~\ref{fig:workflow}, is divided into two main stages: channel selection and auditory attention classification. In the first part, raw EEG signals are transformed into topographic images that encode the alpha-band power distribution across the scalp. These images are then used to train a CNN to classify the direction of auditory attention. Once the CNN is trained, the Shapley value used in \cite{chen2020explaining} is applied to estimate the contribution of each EEG channel to the model’s predictions. Based on these importance scores, the EEG channels are ranked, and the  top-$k$ most informative channels are selected. In the second stage, with the selected channels determined, a TCN is constructed, which directly takes the raw EEG signals as input without image transformation, since time-series data preserve temporal information that is essential for accurate sequence modeling. This TCN is trained and evaluated using only the selected EEG channels, and the performance impact of channel reduction is assessed.
In this section, we first describe the dataset preprocessing, followed by the details of DeepSHAP-guided channel selection and auditory attention detection.

\subsection{Dataset and Preprocessing}

We use a public dataset constructed by the Technical University of Denmark (DTU dataset)~\cite{DTU,fuglsang2017noise, wong2018comparison} to train and test our method. 
The DTU dataset consists of recordings from 18 normal-hearing subjects. In these experiments, subjects were instructed to selectively attend to one of two competing short stories, narrated by one male and one female speaker. EEG signals were recorded using a 64-channel BioSemi system at a sampling rate of 512 Hz. The experiment comprised 60 trials, each lasting 50 seconds, during which the subject attended to one of the two speakers. This results in approximately 50 minutes of EEG data per subject, totaling around 15 hours of EEG recordings across all subjects.

In this work, raw EEG signals were first downsampled to 128 Hz and band-pass filtered between 1 and 45 Hz to remove slow drifts and high-frequency noise, including power-line artifacts. Eye movement artifacts were removed using EOG channels, and all EEG signals were re-referenced to the average of the mastoid electrodes. The continuous EEG data was segmented into overlapping decision windows of 10 seconds, with a 50\% overlap to increase the number of training samples. Finally, each EEG channel was z-score normalized on a per-trial basis to standardize amplitude variability across channels.

\subsection{DeepSHAP-guided EEG Channel Reduction}
\subsubsection{Spectro-Spatial EEG Image Construction}
To extract informative features related to auditory attention, each EEG segment was transformed into a spectro-spatial image representation. Specifically, a Fast Fourier Transform (FFT) was applied to compute the alpha-band power (8–14 Hz) for each channel. The resulting alpha power values were projected onto a 2D plane using the Azimuthal Equidistant Projection, preserving the relative spatial arrangement of the EEG electrodes.

\begin{figure}[tbp]
\centerline{\includegraphics[width=0.9\columnwidth]{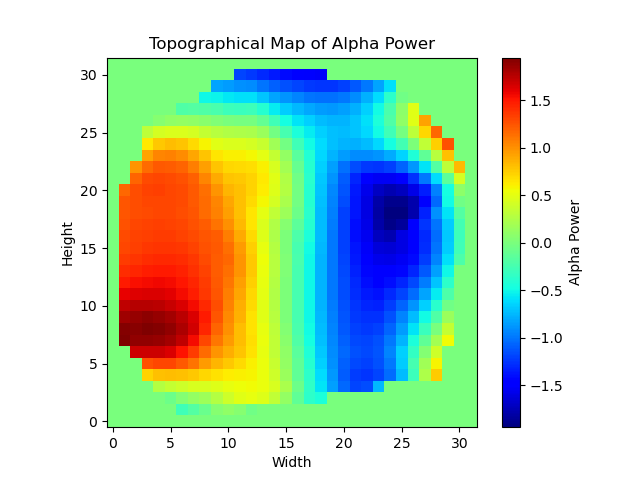}}
\caption{Topographical map of alpha power distribution across the scalp. Warmer colors (red) indicate regions of higher alpha power, while cooler colors (blue) represent lower alpha power. The spatial layout corresponds to the 2D projection of electrode positions on the scalp.}
\label{fig:topo}
\end{figure}

Each EEG channel's alpha power was mapped to its 2D location, and missing pixels were interpolated using cubic interpolation to produce a smooth $32\times32$ topographic image, shown as Fig.~\ref{fig:topo}. This image captures both spectral (alpha power) and spatial (scalp location) information, which are known to be relevant for auditory attention modulation, particularly in the parieto-occipital regions.

\subsubsection{CNN-Based Auditory Attention Classifier}
The topographic EEG images were fed into a CNN to classify the attended direction (left or right ear). The network architecture consists of one convolutional layer followed by three fully connected layers. The input to the model is a single-channel $32\times32$ image representing the spatial distribution of alpha power across the scalp. The convolutional layer applies $32$ filters with a $3\times3$ kernel, a stride of $1$, and padding of $1$, followed by batch normalization, ReLU, and average pooling with a $2\times2$ kernel. 
The resulting feature maps are flattened into a vector of length 8192 and passed through fully connected layers. The final output is a single logit, which is interpreted as a binary classification output indicating the direction of auditory attention. 

\subsubsection{DeepSHAP-Based Channel Importance Estimation}
To identify the most informative EEG channels, we applied DeepSHAP, an explainable AI method that approximates Shapley values~\cite{shapley1953value} using DeepLIFT~\cite{shrikumar2017learning}, to the trained CNN model. A small subset of 200 training samples was used as a background reference. After each cross-validation fold, a 100 test samples of the test set were used to compute the SHAP values for all input pixels. The SHAP values across all samples and cross-validation folds were averaged (absolute mean) to generate a global importance map. The most relevant EEG channels were determined by mapping high-importance pixels back to their corresponding electrode locations.

% To evaluate the impact of channel selection, we retrained and tested the CNN model using only the top k channels (e.g. 16 or 32) and compared the performance to the original full-channel model. This pipeline enabled us to study the trade-off between EEG hardware complexity and classification accuracy.

\subsection{Efficient TCN-based AAD}
Following the channel selection process, we construct a compact TCN to perform AAD using only the top-$k$ selected EEG channels. The goal is to build a low-latency, low-complexity model suitable for efficient real-time inference.
The network consists of two dilated 1D convolutional layers, followed by an adaptive average pooling layer and a lightweight classifier. The first convolutional layer expands the input to 128 feature maps using a kernel size of 7 and a dilation rate of 1. The second layer maintains the same number of channels with a dilation rate of 2, effectively increasing the receptive field without significantly increasing the number of parameters.
Moreover, the model has a negligible memory footprint and requires around 200\,M operations, which makes it well-suited for deployment in wearable EEG-based AAD systems.

The network takes EEG signals in shape $(B, T, C)$, where $B$ is the batch size, $T$ is the number of time steps (e.g., 1280 for a 10-second window at 128 Hz), and $C$ is the number of selected channels. The decision to work directly on sequential EEG signals, rather than image representations used in the selection stage, is intentional: it preserves the temporal resolution of EEG, which is critical for decoding time-varying cognitive states.
Moreover, after channel reduction, image-based representations become less reliable. Interpolating these missing channel values often leads to distorted or spatially unbalanced feature maps, which degrade classification accuracy. In contrast, operating on raw waveforms avoids such interpolation artifacts and allows the model to fully exploit the temporal dynamics of the remaining informative channels.
By integrating channel selection and temporal modeling, this two-stage framework enables accurate and efficient AAD with a reduced number of EEG sensors and model parameters.

\section{Experimental Results}

\subsection{Experimental Setup}
The experiments were conducted separately on the DTU dataset in a subject-wise manner. Random cross-validation was used to split the data into training (80\%), validation (10\%), and test (10\%) sets. The network is trained using the Binary Cross-Entropy with Logits Loss, and optimized with the Adam optimizer using a learning rate of $3\times10^{-4}$ and a weight decay of $3\times10^{-4}$. A stratified split was used to preserve the class distribution across the subsets. 
The data is also inherently equally distributed between the left and right ear. 
Test accuracy was averaged over 10 folds of cross-validation. Early stopping with a patience of 20 epochs was applied to prevent overfitting. 
To evaluate the performance of the channel reduction method, experiments were conducted using decision window lengths of 10 seconds, with reduced channel configurations of 48, 32, 16, and 8 channels. Mean classification accuracy was computed for the original 64-channel setup and compared against the reduced-channel configurations to assess the impact on model performance.

\subsection{Channel Selection}
\begin{figure}[tbp]
\centerline{\includegraphics[width=0.9\columnwidth]{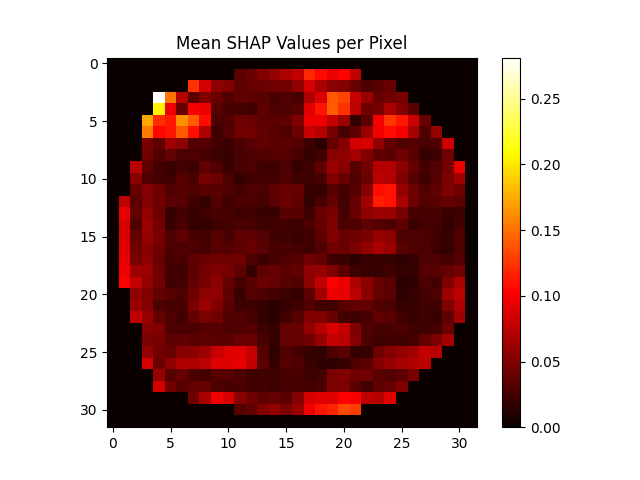}}
\caption{Mean SHAP values for each pixel in all the topographic EEG images of subject 1, indicating their
 relative importance to the model’s prediction. Higher values of SHAP indicate a bigger contribution
 to the model’s decision.}
\label{fig:SHAP}
\end{figure}

\begin{figure*}[htbp]
\centerline{\includegraphics[width=0.85\textwidth]{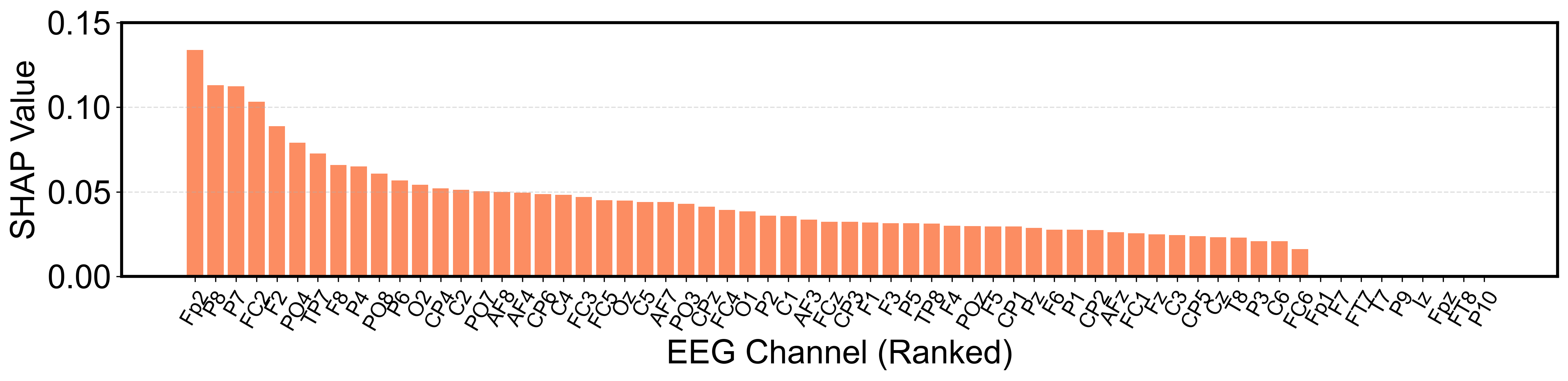}}
\caption{Ranked SHAP value of different EEG channels on Subject 1 (10-s window), ranking in descending order.}
\label{fig:shap_rank}
\end{figure*}

As shown in Fig.~\ref{fig:SHAP}, the averaged SHAP values across all samples and cross-validation folds yield a global importance map, shown here using Subject 1 as an example. Fig.~\ref{fig:shap_rank} illustrates channel-wise SHAP value ranking derived from the trained CNN model. 
Each bar represents the average SHAP value of a specific EEG channel, quantifying its contribution to the model's auditory attention prediction. The channels are arranged in descending order of importance, with higher SHAP values indicating greater influence on the model’s output.  
Notably, only a subset of channels contributes significantly to the decision-making process, for example, Fp2, P8, P7 and FC2 emerge as the most informative channels for auditory attention decoding. While others (e.g. FP1, F7, etc.) show zero SHAP values, suggesting redundancy.

\subsection{AAD Accuracy with Channel Selection}
\begin{figure*}[htbp]
\centerline{\includegraphics[width=0.85\textwidth]{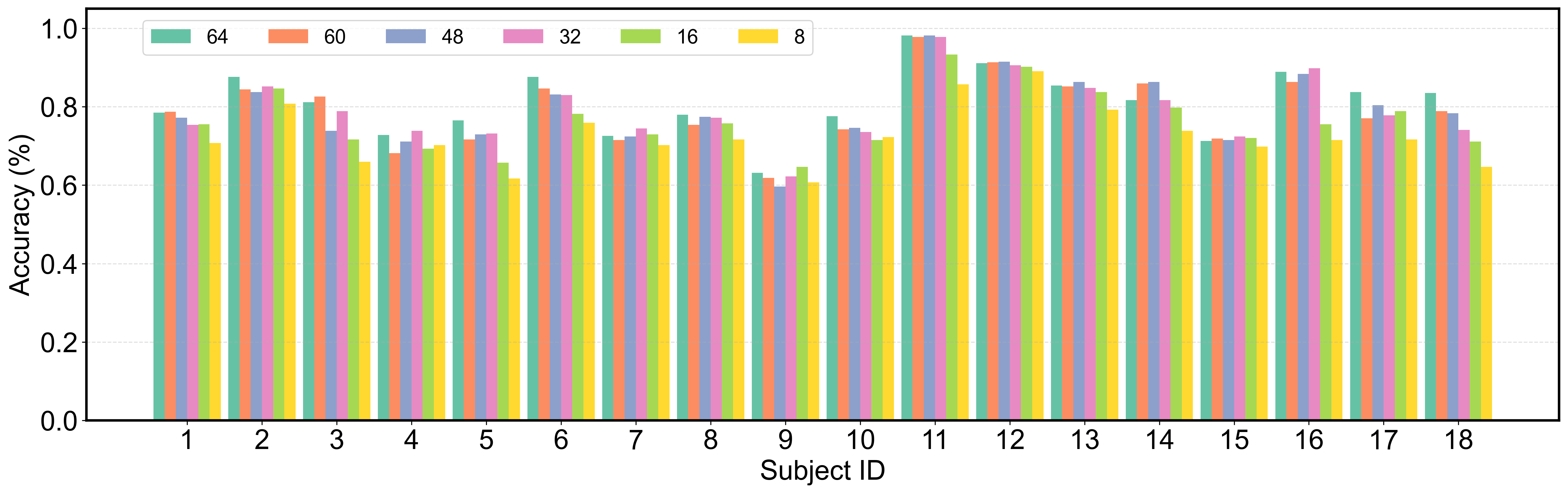}}
\caption{Subject-wise AAD accuracy across the DTU dataset under different EEG channel reductions (10-s window).}
\label{fig:channel_reduction_results}
\end{figure*}

Fig.~\ref{fig:channel_reduction_results} illustrates the subject-wise AAD accuracy across five different EEG channel configurations: 64, 60, 48, 32, 16, and 8 channels, where the 64-channel setup serves as the baseline. Each bar cluster represents the average performance of a single subject under varying levels of channel reduction.

Overall, reducing the number of channels results in a consistent performance drop, but the degree of degradation varies across subjects. In most cases, models using 48 or 32 channels still retain high accuracy levels across nearly all subjects, often within a few percentage points of the full-channel baseline. This demonstrates the effectiveness of the DeepSHAP-guided selection strategy in preserving informative spatial features even under substantial channel reduction. For example, subject 11 exhibits only a marginal difference between the 64- and 32-channel settings (around 0.4\%, 98.15\% vs. 97.78\%). However, reducing to 16 channels results in a more noticeable performance drop to 93.33\%. Similarly, subject 12 still maintains relatively stable performance even under aggressive channel reduction, with accuracy decreasing only slightly from 91.11\% (64 channels) to 89.07\% (8 channels), indicating robustness to information loss. 
Importantly, the performance degradation is not uniform across subjects: while some individuals maintain relatively stable accuracy even at 8 channels, others exhibit a sharper decline. This suggests inter-subject variability in the spatial distribution of attention-related EEG features.

\begin{figure}[htbp]
\centerline{\includegraphics[width=0.7\columnwidth]{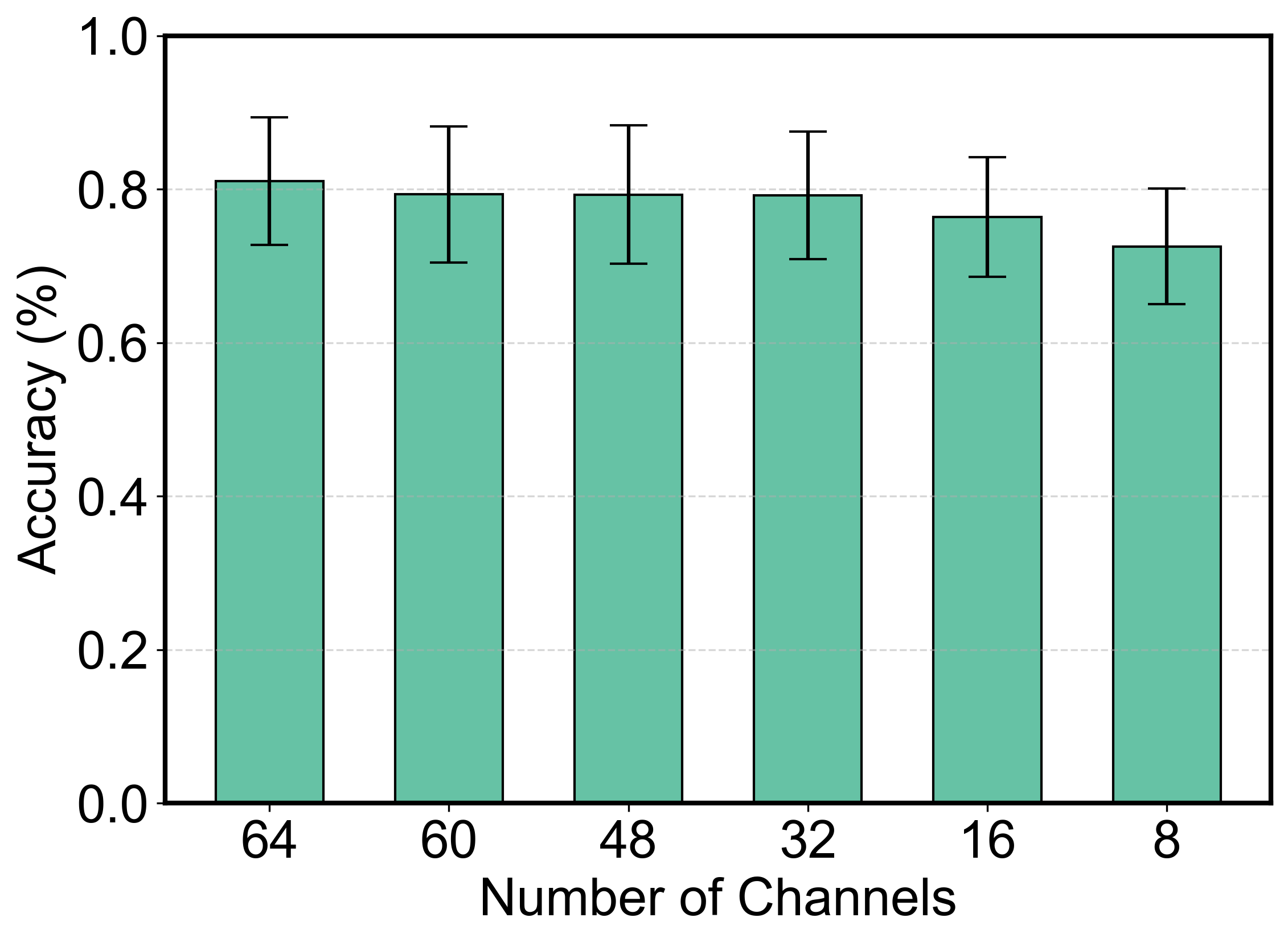}}
\caption{Average AAD accuracy across the DTU dataset under different EEG channel reductions (10-s window).}
\label{fig:overall_results}
\end{figure}

Shown as Fig.~\ref{fig:overall_results}, compared to the 64-channel baseline, the average accuracy across all subjects drops by 1.79\% when reducing to 48 channels, and remains nearly unchanged (0.06\% drop) when further reducing to 32 channels. However, performance decreases become substantial beyond this point, with drops of 2.85\% from 32 to 16 channels, and 3.83\% from 16 to 8 channels. This pattern suggests that moderate reductions (down to 32 channels) preserve most of the discriminative information, while more aggressive reductions begin to compromise model performance.
In general, these results validate the robustness of our approach and highlight its potential for practical AAD systems with reduced electrode setups. Table~\ref{tab:model_complexity} also shows the model size and computational cost under different EEG channel settings.

\begin{table}[htbp]
\centering
\caption{Model size and computational cost under different EEG channel settings (10-s window).}
\begin{tabular}{c|c|c}
\hline
\textbf{\# Channels} & \textbf{Model Size (M)} & \textbf{Computational Cost (MFLOPs)} \\
\hline
\hline
64  & 0.18 & 220 \\
48  & 0.17 & 202 \\
32  & 0.15 & 183 \\
16  & 0.14 & 165 \\
8   & 0.13 & 156 \\
\hline
\hline
\end{tabular}
\label{tab:model_complexity}
\end{table}

\section{Conclusion}
In this paper, we proposed SHAP-AAD, a two-stage framework that combines DeepSHAP-based channel selection with an efficient TCN for auditory attention detection using pure EEG signals. Our method effectively reduces the number of EEG channels while preserving high classification accuracy. Experimental results demonstrate that models using only 32 or even 8 selected channels for some specific cases can achieve performance comparable to the full 64-channel setup, validating the robustness and efficiency of our approach.

\newpage
\bibliographystyle{IEEEtran}
\bibliography{IEEEabrv,ref} 

\end{document}